\documentclass[11pt,fleqn]{article}
\usepackage[cp1251]{inputenc}
\usepackage[english]{babel}
\usepackage{amsmath,amssymb,euscript,amsthm,amsfonts}
\newtheorem{thm}{Theorem}
\theoremstyle{definition}

\theoremstyle{remark}
\newtheorem{remk}{Remark}

\begin{document}
\allowdisplaybreaks

\renewcommand{\proofname}{Proof}

\makeatletter \headsep 10 mm \footskip 10 mm
\renewcommand{\@evenhead}%
{\vbox{\hbox to\textwidth{\strut \centerline{{\it Maryna
Nesterenko, Poman Popovych}}} \hrule}}

\renewcommand{\@oddhead}%
{\vbox{\hbox to\textwidth{\strut \centerline{{\it Realizations of real semisimple low-dimensional Lie algebras
}}} \hrule}}
\begin{center}
{\Large\bf Realizations of real semisimple \\low-dimensional Lie algebras }
\end{center}

\begin{flushleft} 
Maryna O. NESTERENKO
Roman O. POPOVYCH,
\end{flushleft}

\noindent Institute of Mathematics of NAS of Ukraine, 3
Tereshchenkivska Str., Kyiv 4, 01601 Ukraine\\
E-mail: maryna@imath.kiev.ua, rop@imath.kiev.ua\\
URL: http://www.imath.kiev.ua/\~{}maryna/,\ 
http://www.imath.kiev.ua/\~{}rop/

\begin{abstract}
A complete set of inequivalent realizations of three- and four-dimen\-sion\-al
real unsolvable Lie algebras in vector fields on a space of an arbitrary (finite)
number of variables is obtained.
\end{abstract}

Representations of Lie algebras by vector fields are widely
applicable e.g. in integrating of ordinary differential equations,
group classification of partial differential equations, the theory
of differential invariants, general relativity and other physical
problems.
There exist many papers devoted to the problem of construction of
realizations of Lie algebras. All possible realizations of Lie algebras in
vector fields on the two-dimensional complex and real spaces were first classified
by S.Lie himself~[\ref{lie1880}, \ref{lie1893}].

In this paper  we construct a complete set of inequivalent faithful realizations of unsolvable real
Lie algebras of dimension no greater than four in vector fields on
a space of an arbitrary (finite) number of variables.

A necessary step to classify realizations of Lie algebras is
classification of these algebras, i.e.\ classification of possible
commutative relations between basis elements. Complete classification of
Lie algebras of dimension up to and including six
can be found in the papers of V.V.~Morozov [\ref{Morozov}],
G.M.~Mu\-ba\-rak\-zya\-nov [\ref{Mubarakzyanov1},
\ref{Mubarakzyanov2}, \ref{Mubarakzyanov3}] and P.~Turkowski
[\ref{Turkowski2}]. The problem of classification of Lie algebras of higher
orders is solved only for some classes e.g.\ the simple and semi-simple algebras.

There are exist four unsolvable real Lie algebras of dimension no greater than four
(here $q=1,2,3$):
\[
\begin{array}{llll}
sl(2,\mathbb{R}){:}&[e_1,e_2]=e_1,&[e_1,e_3]=2e_2,&[e_2,e_3]=e_3;\\[1ex]
so(3){:}&[e_1,e_2]=e_3,&[e_3,e_1]=e_2,&[e_2,e_3]=e_1;\\[1ex]
sl(2,\mathbb{R})\oplus A_1{:}&[e_1,e_2]=e_1,&[e_1,e_3]=2e_2,&[e_2,e_3]=e_3, \;\;[e_q,e_4]=0;\\[1ex]
so(3)\oplus A_1{:}&[e_1,e_2]=e_3,&[e_3,e_1]=e_2,&[e_2,e_3]=e_1, \;\;[e_q,e_4]=0.
\end{array}
\]

\begin{remk} {\it Notations and conventions.}
Below $\partial_a=\partial / \partial x_a$,
$x=(x_1,\ldots, x_n)$, $\check x=(x_3,\dots, x_n)$, $\hat x=(x_4, \dots, x_n)$,
$a=\overline{1,n}$, $j,k=\overline{4,n}$.
We use convention on summation over repeat indexes.
We denote the $N$-th realization of an algebra $A$ as $R(A,N)$.
\end{remk}

To classify realizations of a $m$-dimensional Lie algebra $A$ in
the most direct way, we have to take $m$ linearly independent
vector fields of the general form
$e_s=\xi^{sa}(x)\partial_a,$ $s=\overline{1,m}$,
and require them to satisfy the commutation relations of $A.$
As a result, we obtain a system of first-order PDEs for the
coefficients~$\xi^{sa}$ and then we integrate it, considering all
the possible cases. For each case we transform the solution into
the simplest form, using either local diffeomorphisms of the
space of $x$ and automorphisms of $A$ if we looking for the weakly
inequivalent classification or only local diffeomorphisms of the
space of $x$ if the strong inequivalence is meant. A~disadvantage
of this method is the necessity to solve a complicated nonlinear
system of PDEs. Another way is to classify sequentially
realizations of a series of nested subalgebras of $A,$ starting
with a one-dimensional subalgebra or other subalgebra with known realizations
and ending up with $A.$
Thus, to prove the following theorem, we apply the above method,
starting from the algebra $A_{2.1}$ formed by $e_1$ and $e_2$.

\begin{thm}
Let first-order differential operators satisfy the commutation
relations of $sl(2,{\mathbb R})$. Then there exist transformations
reducing these operators to one of the forms:

\vspace{1ex}

1) $\partial_1$, $x_1\partial_1+x_2\partial_2$, $x_1^2\partial_1+2x_1x_2\partial_2+x_2\partial_3;$

\vspace{0.5ex}

2) $\partial_1$, $x_1\partial_1 +x_2\partial_2$, $(x_1^2+x_2^2)\partial_1+2x_1x_2\partial_2$;

\vspace{0.5ex}

3) $\partial_1$, $x_1\partial_1 +x_2\partial_2$, $(x_1^2-x_2^2)\partial_1+2x_1x_2\partial_2$;

\vspace{0.5ex}

4) $\partial_1$, $x_1\partial_1 +x_2\partial_2$, $x_1^2\partial_1+2x_1x_2\partial_2$;

\vspace{0.5ex}

5) $\partial_1$, $x_1\partial_1$, $x_1^2\partial_1$.
\end{thm}

\begin{thm}
A complete list of inequivalent realizations of $sl(2,{\mathbb R})\oplus A_1$
is exhausted by the following ones:

\vspace{1 ex}

1) $\partial_1, \ x_1\partial_1+x_2\partial_2,\
x_1^2\partial_1+2x_1x_2\partial_2+x_2\partial_3, \partial_4$;

\vspace{0.5 ex}

2) $\partial_1, x_1\partial_1 +x_2\partial_2,
x_1^2\partial_1+2x_1x_2\partial_2+x_2\partial_3,
x_2\partial_1+2x_2x_3\partial_2+(x_3^2+x_4)\partial_3$;

\vspace{0.5 ex}

3) $\partial_1$, $x_1\partial_1 +x_2\partial_2$,
$x_1^2\partial_1+2x_1x_2\partial_2+x_2\partial_3$,
$x_2\partial_1+2x_2x_3\partial_2+(x_3^2+c)\partial_3$,
\hspace*{9.5mm}$c\in \{-1;0;1\}$;

\vspace{0.5 ex}

4) $\partial_1$, $x_1\partial_1 +x_2\partial_2$, $(x_1^2+x_2^2)\partial_1+2x_1x_2\partial_2$, $\partial_3$;

\vspace{0.5 ex}

5) $\partial_1$, $x_1\partial_1 +x_2\partial_2$, $(x_1^2-x_2^2)\partial_1+2x_1x_2\partial_2$, $\partial_3$;

\vspace{0.5 ex}

6) $\partial_1$, $x_1\partial_1 +x_2\partial_2$, $x_1^2\partial_1+2x_1x_2\partial_2$, $\partial_3$;

\vspace{0.5 ex}

7) $\partial_1$, $x_1\partial_1 +x_2\partial_2$, $x_1^2\partial_1+2x_1x_2\partial_2$, $x_2x_3\partial_2$;

\vspace{0.5 ex}

8) $\partial_1$, $x_1\partial_1 +x_2\partial_2$, $x_1^2\partial_1+2x_1x_2\partial_2$, $x_2\partial_2$;

\vspace{0.5 ex}

9) $\partial_1$, $x_1\partial_1$, $x_1^2\partial_1$, $\partial_2$.
\end{thm}

\begin{proof}
The automorphism group of the algebra $sl(2,\mathbb{R})\oplus A_1$
is a direct product of the automorphism groups of $sl(2,\mathbb{R})$ and $A_1$.
We extend the realizations of $sl(2,\mathbb{R})$ to realizations of  $sl(2,\mathbb{R})\oplus A_1$
with the operator $e_4$, beginning from the most general form
$e_4=\eta^a(x)\partial_a.$

Consider the realization $R(sl(2,\mathbb{R}),1)$.\
The general form of the operator $e_4$ which commutates with the basis elements
of $R(sl(2,\mathbb{R}),1)$ is as follows:
\[
e_4=\xi^1x_2\partial_1+ (2\xi^1x_3+\xi^2)x_2\partial_2+(\xi^1x_3^2+\xi^2x_3+\xi^3)\partial_3+\xi^j\partial_j,
\]
where $\xi^a$ are arbitrary functions of $\hat x$.
The  form of operators $e_1$, $e_2$ and $e_3$ is preserved by the transformation:
\[
\tilde{x}_1=x_1+\frac{f^1x_2}{1-f^1x_3},\quad \tilde{x}_2=\frac{f^2x_2}{(1-f^1x_3)^2},
\quad
\tilde{x}_3=\frac{f^2x_3}{1-f^1x_3}+f^3,\quad \tilde{x}_j=f^j,
\]
where $f^a$ are arbitrary functions of $\hat x$.
After action of this transformation the operator $e_4$ turns
into the operator $\tilde e_4$ of the same form with following functions
$\tilde{\xi^a}$:
\begin{gather*}
\tilde{\xi}^1=\frac{1}{f^2}(\xi^1+\xi^2f^1+\xi^3(f^1)^2+\xi^jf^1_j),\quad
\tilde{\xi}^2=\xi^2+2\xi^3f^1-2\tilde{\xi}^1(f^3)^2+\xi^j\frac{f^2_j}{f^2},
\\
\tilde\xi^3=\xi^3f^2-\tilde\xi^1(f^3)^2-\tilde\xi^2f^3+\xi^3f^2+\xi^jf^3_j,\quad
\tilde\xi^j=\xi^kf^j_k.
\end{gather*}
Here and below subscripts mean differentiation with respect to the corresponding variables $x_a$.

There are two possible cases.

1) $\exists j{:}\;\xi^j\ne 0$. Then the operator $e_4$ can be transformed to the form $\tilde
e_4=\partial_4$ and we obtain the realization $R(sl(2,\mathbb{R})\oplus A_1,1)$.

2) $\tilde{\xi}^j=0$.
The expression $I=(\xi^2)^2-4\xi^1\xi^3$ is an invariant of the above transformations of $\xi$.
Therefore, we can make $ \tilde{\xi}_1=1$, $\tilde{\xi}_2=0,$ $\tilde{\xi}_3=I$.
If $I=\mathrm{const}$ then we obtain the realization $R(sl(2,\mathbb{R})\oplus A_1,3)$,
otherwise we can choose new variable $\tilde x_4=I$ and obtain
the realization $R(sl(2,\mathbb{R})\oplus A_1,2)$.

We omit calculations on the realizations
$R(sl(2,\mathbb{R})\oplus A_1,4\mbox{--}9)$,
because they are simpler than the adduced ones and are made in the same way, starting from
three other realizations of the algebra $sl(2,\mathbb{R})$.

Inequivalence of the obtained realizations can be easily proved by means
of technics proposed in~[\ref{Nesterenko1}].
\end{proof}

\begin{thm}
There are only two inequivalent realizations of the algebra $so(3)$:

\vspace{1 ex}

1) $-\sin x_1 \tan x_2\partial_1-\cos x_1\partial_2$,
$-\cos x_1\tan x_2 \partial_1 +\sin x_1 \partial_2$, $\partial_1$;

 \vspace{0.5 ex}

2) $-\sin x_1 \tan x_2 \partial_1-\cos x_1\partial_2+\sin x_1\sec x_2\partial_3$,

$\phantom{2)\, }-\cos x_1\tan x_2 \partial_1 +\sin x_1 \partial_2+\cos x_1\sec x_2\partial_3$, $\partial_1$.
\end{thm}
\begin{remk}
The realizations $R(so(3),1)$ and $R(so(3),2)$ are well-known.
At the best of our knowledge, completeness of the list of these realizations
was first proved in~[\ref{zhdanov2000}].
We do not assert that the adduced forms of realizations are optimal for all applications
and the classification from Theorem 3 is canonical.
Consider the realization $R(so(3),1)$ of rank 2 in more details.
It acts transitively on the manifold $S^2$.
With the stereographic projection $\tan{x_1}=t/x$, ${\rm cotan}\ x_2=\sqrt{x^2+t^2}$
it can be reduced to the well known realization on the plane~[\ref{Olver}]:
\[
(1+t^2)\partial_t+xt\partial_x,\quad\ x\partial_t-t\partial_x, \quad
-xt\partial_t-(1+x^2)\partial_x
\]
If dimension of the $x$-space is not smaller than $3$, the variables $x_1$, $x_2$ and
the implicit variable $x_3$ in $R(so(3),1)$ can be interpreted as the angles and the radius
of the spherical coordinates (imbedding $S^2$ in ${\mathbb R}^3$).
Then in the corresponding Cartesian coordinates $R(so(3),1)$ has the well-known form:
\[
x_2\partial_3-x_3\partial_2,\qquad x_3\partial_1-x_1\partial_3,\qquad x_1\partial_2-x_2\partial_1,
\]
which is generated by the standard representation of $SO(3)$ in ${\mathbb R}^3$.
\end{remk}

\begin{thm}
A list of inequivalent realizations of the algebra $so(3)\oplus A_1$
in vector fields on a space of an arbitrary (finite) number of variables is exhausted by
the following ones:

\vspace{1 ex}

1) $-\sin x_1 \tan x_2\partial_1-\cos x_1\partial_2$,
$-\cos x_1\tan x_2 \partial_1 +\sin x_1 \partial_2$, $\partial_1$, $\partial_3$;

\vspace{0.5 ex}

2) $-\sin x_1 \tan x_2 \partial_1-\cos x_1\partial_2+\sin x_1\sec
x_2\partial_3$,

$\phantom{2)}$~$-\cos x_1\tan x_2 \partial_1 +\sin x_1
\partial_2+\cos x_1\sec x_2\partial_3$,
$\partial_1$, $\partial_3$;

\vspace{0.5 ex}

3) $-\sin x_1 \tan x_2 \partial_1-\cos x_1\partial_2+\sin x_1\sec x_2\partial_3$,

$\phantom{3)}$~$-\cos x_1\tan x_2 \partial_1 +\sin x_1
\partial_2+\cos x_1\sec x_2\partial_3$,
$\partial_1$, $x_4\partial_3$;

\vspace{0.5 ex}

4) $-\sin x_1 \tan x_2 \partial_1-\cos x_1\partial_2+\sin x_1\sec
x_2\partial_3$,

$\phantom{4)}$~$-\cos x_1\tan x_2 \partial_1 +\sin x_1
\partial_2+\cos x_1\sec x_2\partial_3$, $\partial_1$,
$\partial_4$.
\end{thm}

\begin{proof}
The automorphism group of $so(3)\oplus A_1$
is the direct product of the automorphism groups of $so(3)$ and $A_1$.
To classify realizations of $so(3)\oplus A_1$,
we start from the realizations $R(so(3),1)$ and $R(so(3),2)$.
For convenience we rewrite them as a realization parameterized with $\alpha \in\{0;1\}$:
\begin{gather*}
e_1=-\sin x_1 \tan x_2 \partial_1-\cos x_1\partial_2+\alpha \sin x_1\sec x_2\partial_3,\\
e_2=-\cos x_1\tan x_2 \partial_1 +\sin x_1 \partial_2+\alpha \cos x_1\sec x_2\partial_3,\\
e_3=\partial_1
\end{gather*}
The values $\alpha=0$ and $\alpha=1$ correspond to the realizations $R(so(3),1)$ and $R(so(3),2)$.

We take the operator $e_4$ in the most general form $e_4=\xi^a(x)\partial_a$
and obtain the equations for $\xi^a(x)$ from condition of vanishing commutators of $e_4$ with
the other basis elements:
\begin{subequations}
\begin{gather}
\xi^a_1=0,\quad \xi^2_2=0,\quad \xi^j_2=0,\label{1a}\\
\alpha\xi^2_3-\xi^1\cos x_2=0,\quad
\alpha\xi^1_3\cos x_2+\xi^2=0,\quad
\xi^3_2\cos x_2+\alpha\xi^1=0,\label{1b}\\
\xi^1_2-\xi^1\tan x_2 =0,\quad
\alpha\xi^3_3-\alpha\xi^2\tan x_2=0, \quad
\alpha\xi^j_3=0.\label{1c}
\end{gather}
\end{subequations}

It follows from (\ref{1a}) that $\xi^2=\xi^2(\check x)$ and $\xi^j=\xi^j(\check x)$.

In the case $\alpha=0$ we obtain from (\ref{1b}) that
$\xi^1=0$, $\xi^2=0$ and $\xi^3=\xi^3(\check x)$.
Then $e_4$ has the form $e_4=\xi^p(\check x)\partial_p$, where $p=\overline{3,n}$, and
one of the coefficients $\xi^p$ does not vanish.
Using allowable transformations of variables
$\tilde x_1=x_1,$ $\tilde x_2=x_2,$ $\tilde x_p=f^p(\check x)$,
we can can make $\xi^3=1$ and $\xi^j=0$.
("Allowable" means that such transformations preserve the form of $e_1$, $e_2$ and $e_3$.)
As a result, we obtain realization $R(so(3)\oplus A_1,1)$.

Consider the case $\alpha=1$.
The general solution of system (\ref{1a})--(\ref{1c}) is:
\begin{gather*}
\xi^1=\frac{\varphi^1\sin x_3+\varphi^2\cos x_3}{\cos x_2},\quad
\xi^2=\varphi^2\sin x_3-\varphi^1\cos x_3,\\
\xi_3=\varphi^3-(\varphi^1\sin x_3+\varphi^2\cos x_3)\tan x_2,\quad \xi^j=\varphi^j,
\end{gather*}
where $\varphi^a$ are arbitrary functions of $\hat x$.
Therefore, the operator $e_4$ can be presented in the form:
\[
e_4=\varphi^1e_1'+\varphi^2e_2'+\varphi^3e_3'+\varphi^j\partial_j,
\]
where operators $e'_1$--$e'_3$ are obtained from $e_1$--$e_3$
with transposition of the variables $x_1$ and $x_3$.

The next step is to simplify the operator $e_4$.
Since in this case the direct method of finding allowable transformations of variables
is too cumbersome and complicated, we use the infinitesimal approach.

An one-parametric group of local transformations in the space of variables $x$ preserves the form of
operators $e_1$--$e_3$ if its infinitesimal generator $Q$ commutes with these operators.
Therefore, $Q$ has the same form as $e_4$:
\[
Q=\rho^1e_1'+\rho^2e_2'+\rho^3e_3'+\rho^j\partial_j,
\]
where $\rho^a$ are arbitrary functions of $\hat x$.

There are two possible cases: $\xi^j=0$ or $\exists j{:}\ \xi^j\ne 0$.\
In any case the operator~$e_4$ can be transformed
by means of allowable transformations
$\tilde x_1=x_1$, $\tilde x_2=x_2$, $\tilde x_3=x_3$, $\tilde x_j=f^j(\hat x)$
to the form:
\[
e_4=\varphi^1e_1'+\varphi^2e_2'+\varphi^3e_3'+\beta\partial_4,\quad \beta\in \{0,1\}.
\]
Below we use only transformations preserving $\hat x$ and, therefore, assume $\rho^j=0$.

Introducing the vector notations $\bar \varphi=(\varphi^1,\varphi^2,\varphi^3)$,
$\bar\rho=(\rho^1,\rho^2,\rho^3)$, $\bar\rho_4=(\rho^1_4,\rho^2_4,\rho^3_4)$, and
$\bar e'=(e_1',e_2',e_3')$, we can present the commutator $[e_4,Q]$ as follows:
\[
[e_4,Q]=(\bar\rho\times \bar \varphi-\beta\bar{\rho}_4)\cdot\bar e',
\]
where "$\times$" and "$\cdot$" denote the vector and scalar products.
The finite transformations $\widetilde{\bar{\varphi}}=\bar\gamma(\varepsilon,\bar\varphi,\hat x)$
generated by $Q$ are found by integrating the Lie equations:
\begin{gather}\label{slefaoso}
\frac{d\bar\gamma}{d\varepsilon}=\bar\rho\times \bar \gamma-
\beta\bar\rho_4,\quad \bar\gamma |_{\varepsilon=0}=\bar \varphi,
\end{gather}
where $\varepsilon$ is a group parameter and $\hat x$ are assumed constants.
Therefore,
\[
\bar \gamma=OJ(\varepsilon)O^{\rm T}\bar\varphi-
\beta O\!\int_0^{\varepsilon}\!\!J(\varepsilon)d\varepsilon\; O^{\rm T}\bar\rho_4,
\]
where $O$ is an orthogonal matrix having $\bar\rho/|\bar\rho|$ as the third column,
\[
J(\varepsilon)=\left(
\begin{array}{ccc}
\cos |\bar\rho|\varepsilon & -\sin |\bar\rho|\varepsilon & 0\\
\sin |\bar\rho|\varepsilon & \cos |\bar\rho|\varepsilon & 0\\
0&0&1
\end{array}
\right).
\]

In the case $\beta=0$ $\varphi^1$ and $\varphi^2$ can be made to vanish by means of
transformations $\widetilde{\bar{\varphi}}=\bar\gamma(\varepsilon,\bar\varphi,\hat x)$
hence $\tilde e_4=\tilde \varphi^3(\hat x)\partial_3$.
As a result we obtain the realizations $R(so(3)\oplus A_1,2)$ and $R(so(3)\oplus A_1,3)$
if $\varphi_3={\rm const}$ or $\varphi_3\not={\rm const}$ correspondingly.

In the case $\beta=1$ we choose $\bar{\rho}$ as a solution of the system
\begin{equation}\label{sfrho}
\int_0^{\varepsilon}\!\!J(\varepsilon)d\varepsilon\; O^{\rm T}\bar\rho_4=
J(\varepsilon)O^{\rm T}\bar\varphi,
\end{equation}
where $\varepsilon$ is fixed such that there exist
$(\int_0^{\varepsilon}\!\!J(\varepsilon)d\varepsilon)^{-1}$.
Then the transformed $\bar\varphi$ vanishes, i.e. $\tilde e_4=\partial_4$ and we have
the realization $R(so(3)\oplus A_1,4)$.
\end{proof}

The complete classification of realizations of all Lie algebras of
dimension up to and including 4 is adduced in~[\ref{Nesterenko1}].

\vspace{1ex}

\noindent
{\bf Acknowledgments.} The authors are grateful to Profs.\ V.~Boy\-ko,  A.~Ni\-ki\-tin and
I.~Yehorchenko for useful discussions.
The research of MN was supported by National Academy of Science of Ukraine
in the form of the grant for young scientists.

\newpage

\vskip15 pt
\centerline{\Large REFERENCES} \vskip 7 pt
\begin{enumerate}

\small

\item
\label{lie1880}
Lie S. (1880), {\it Theorie der Transformationsgruppen},
Math. Ann., {\bf V.16}, 441--528.

\item
\label{lie1893}
Lie S. (1893), {\it Classification und Integration von gew\"ohnlichen
Differentialgleichungen zwischen $x$, $y$, die eine Gruppe von Transformationen gestatten},
Arch. Math. Naturv., {\bf V.9}, 371--393.

\item
\label{Morozov}
Morozov V.V. (1958), {\it Classification of nilpotent Lie algebras of sixth order},
Izv. Vys. Ucheb. Zaved. Matematika, {\bf N~4 (5)}, 161--171.

\item
\label{Mubarakzyanov1}
Mubarakzyanov~G.M. (1963), {\it On solvable Lie algebras},
Izv. Vys. Ucheb. Zaved. Matematika, {\bf N~1 (32)}, 114--123.

\item
\label{Mubarakzyanov2}
Mubarakzyanov~G.M. (1963), {\it The classification of the real
structure of five-dimensional Lie algebras},
Izv. Vys. Ucheb. Zaved. Matematika, {\bf N~3 (34)}, 99--106.

\item
\label{Mubarakzyanov3}
Mubarakzyanov~G.M. (1963), {\it Classification of solvable Lie
algebras of sixth order with a non-nilpotent basis element},
Izv. Vys. Ucheb. Zaved. Matematika, {\bf N~4 (35)}, 104--116.

\item
\label{Turkowski2}
Turkowski~P. (1990), {\it Solvable Lie algebras of dimension six},
J. Math. Phys., {\bf V.31}, 1344--1350.

\item
\label{zhdanov2000}
Zhdanov R.Z., Lahno V.I. and Fushchych W.I. (2000),
{\it On covariant realizations of the Euclid group},
Comm. Math. Phys., {\bf V.212}, 535--556.

\item
\label{Nesterenko1} Popovych R.O., Boyko V.M., Nesterenko M.O.,
Lutfullin M.W. (2003), {\it Realizations of real low-dimensional
Lie algebras}, J. Phys. A, {\bf 36}, 7337--7360.

\item
\label{Olver}
 Gonzalez-Lopez A., Kamran N. and Olver P.J. {1992} {\it Lie algebras of vector fields in the real plane},
 Proc. London Math. Soc., V.64, 339--368.

\end{enumerate}

\end{document}